\documentclass[aps,prb,superscriptaddress,showpacs,floatfix,twocolumn]{revtex4}

\usepackage[dvips]{graphics}
\usepackage{graphicx}
\usepackage{psfrag}

\renewcommand{\u}{\underline}
\renewcommand{\b}{\mathbf}

\renewcommand{\hat}{\widehat}

\begin{document}

\title{Effects of short-range order on the electronic structure of disordered metallic systems}

\author{Derwyn A. Rowlands}
\affiliation{H.H.~Wills Physics Laboratory, University of Bristol, Bristol BS8 1TL, U.K.}
\author{Julie B. Staunton}
\affiliation{Dept.~of Physics, University of Warwick, Coventry CV4 7AL, U.K.}
\author{Balazs L. Gy\"{o}rffy}
\affiliation{H.H.~Wills Physics Laboratory, University of Bristol, Bristol BS8 1TL, U.K.}
\author{Ezio Bruno}
\affiliation{Dipartimento di Fisica, Universita di Messina Salita Sperone 31, 98166 Messina, Italy}
\author{Beniamino Ginatempo}
\affiliation{Dipartimento di Fisica, Universita di Messina Salita Sperone 31, 98166 Messina, Italy}

\date{November 12, 2004}

\begin{abstract}
For many years the Korringa-Kohn-Rostoker coherent-potential approximation (KKR-CPA) has been widely used to describe the electronic structure of disordered systems based upon
a first-principles description of the crystal potential. However, as a single-site theory the KKR-CPA is unable to account for important environmental effects such as
short-range order (SRO) in alloys and spin fluctuations in magnets, amongst others. Using the recently devised KKR-NLCPA (where NL stands for nonlocal), we show how to remedy
this by presenting explicit calculations for the effects of SRO on the electronic structure of the $bcc$ $Cu_{50}Zn_{50}$ solid solution.
\end{abstract}

\pacs{71.15.Ap, 71.23.-k, 71.20.Be, 71.15.Mb} \maketitle

\section{Introduction}

Currently, the first-principles theory of electrons in disordered metals is based upon Density Functional Theory (DFT) and either the Korringa-Kohn-Rostoker Coherent-Potential
Approximation (KKR-CPA) \cite{Gyorffy1,Stocks1,Gyorffy2} or its stripped down version, the LMTO-CPA, \cite{Kudrnovsky1} for averaging over the random configurations. This
approach has been successfully applied to cases where the disorder is internal as well as external. Examples of the latter are metallic solid solutions such as $Cu_cZn_{(1-c)}$
and $Cu_cPd_{(1-c)}$ above their ordering temperatures $T_0$. Examples of the former are $Fe$ or $Ni$ above their Curie temperatures $T_c$, where randomness in the crystal
potential seen by an electron is the consequence of Disordered Local Moments (DLM), \cite{Staunton2} and secondly valence fluctuating systems such as $Ce$. \cite{Luders1}
Despite significant achievements, \cite{Faulkner2,Gyorffy7} this methodology suffers from the shortcoming of not describing correlations in the fluctuations of the crystal
potential. However a generalization of the KKR-CPA theory has recently been proposed, the KKR-NLCPA \cite{Rowlands1,Rowlands2} (where NL stands for nonlocal), which
systematically takes into account such correlations, enabling environmental effects such as short-range order (SRO) to be taken into account. Although the full 3D KKR-NLCPA
formalism was given in Refs.~\onlinecite{Rowlands1,Rowlands2}, in this paper we present the first realistic 3D implementation of the theory \cite{Rowlands3} by illustrating the
effects of SRO on the $Cu_{50}Zn_{50}$ system.

The physics of the above SRO plays a particularly important role near phase transitions where it is frequently a precursor for long range order and can be said to be driving
the ordering process. For example, in the ordering of the $Cu_{50}Zn_{50}$ solid solution into an intermetallic compound of $B2$ symmetry the system lowers its free energy by
having unlike neighbours more frequently than like neighbours even in the disordered state, thereby lowering the temperature $T_0$ where the system must finally order. Such SRO
is also central to the understanding of electronic transport in general and in $K$-state alloys in particular. \cite{Thomas2,Nicholson1} Moreover, the formation of the moment
in the DLM state of $Ni$~\cite{Staunton2} and the creation of $\gamma$-like $Ce$ atoms near the $\gamma-\alpha$ transition~\cite{Luders1} will be materially affected by the
SRO. The KKR-NLCPA method to be illustrated here will enable these important problems to be tackled in a parameter-independent and material-specific way. However, before
getting on with the task we comment briefly on efforts addressing the same problem as we do by adopting alternative strategies.

There have been several attempts to develop cluster generalisations of the single-site CPA. As with the CPA, the main construct is an effective medium so that the motion of an
electron through it approximates the motion, {\it on the average}, of the electron in the disordered system. An early example, the Molecular CPA (MCPA), \cite{Tsukada1}
introduces a supercell so that the medium has the unsatisfactory attribute of broken translational symmetry. The Embedded Cluster Method (ECM) \cite{Gonis2,Gonis1} refers to
the non-self consistent embedding of a cluster with all the relevant disorder configurations into the CPA medium. The Travelling Cluster Approximation (TCA) \cite{Mills1} based
on diagrammatic methods and the Cluster-CPA (C-CPA) \cite{Mookerjee1,Razee1} based on the Augmented Space Formalism (ASF) \cite{Mookerjee2,Kaplan2} are, like the NLCPA,
satisfactory on account of their translationally invariant, self-consistently determined effective media and herglotz analytic properties. Both the TCA and C-CPA become rapidly
computationally intractable, however, and a KKR version of the latter has been applied only to model systems. \cite{Razee1,Rajput1} A reasonably good alternative starting point
for the electronic structure of some disordered alloys is the tight-binding (TB)-LMTO method \cite{Kudrnovsky1} combined with the CPA, which can include an approximate
treatment of the charge self-consistency needed for a DFT. Mookerjee and Prasad \cite{Mookerjee3} have developed a generalised ASF with correlated variables to describe SRO,
which has been combined with the TB-LMTO and real space recursion technique. \cite{Haydock1} This approach has been used successfully to describe effects of SRO on the
densities of states of several alloy systems. \cite{Saha1,Saha2} Nonetheless it is desirable to develop a computationally tractable generalisation of the CPA within the KKR
method, the KKR-NLCPA, with fewer approximations and superiority with regards to accuracy and reliability over LMTO methods. It will also be amenable for incorporation into a
full DFT description of disordered materials with SRO.\cite{Rowlands4}

This paper is organised as follows. In Sec.~II we briefly summarise the idea of the KKR-NLCPA (for the full derivation see Refs.~\onlinecite{Rowlands1,Rowlands2}), and in
particular we clearly explain how to carry out the fundamental `coarse-graining' procedure for general lattices. Our aim is to show how current KKR-based computational codes
can be straightforwardly adapted to include the KKR-NLCPA with its capability of dealing with disordered systems with SRO. In Sec.~III we present results including SRO
calculations for the $Cu_{50}Zn_{50}$ system, and we conclude in Sec.~IV.

\section{Formalism}

\begin{figure}[!]
 \begin{center}
 \psfrag{Real Space}[][B1][2][0]{Real Space}
 \psfrag{Reciprocal Space}[][B1][2][0]{Reciprocal Space}
 \psfrag{(a)}[][B1][2][0]{(a)}
 \psfrag{(b)}[][B1][2][0]{(b)}
 \psfrag{(c)}[][B1][2][0]{(c)}
 \psfrag{(d)}[][B1][2][0]{(d)}
 \psfrag{(e)}[][B1][2][0]{(e)}
 \psfrag{(f)}[][B1][2][0]{(f)}
 \scalebox{0.55}{\includegraphics{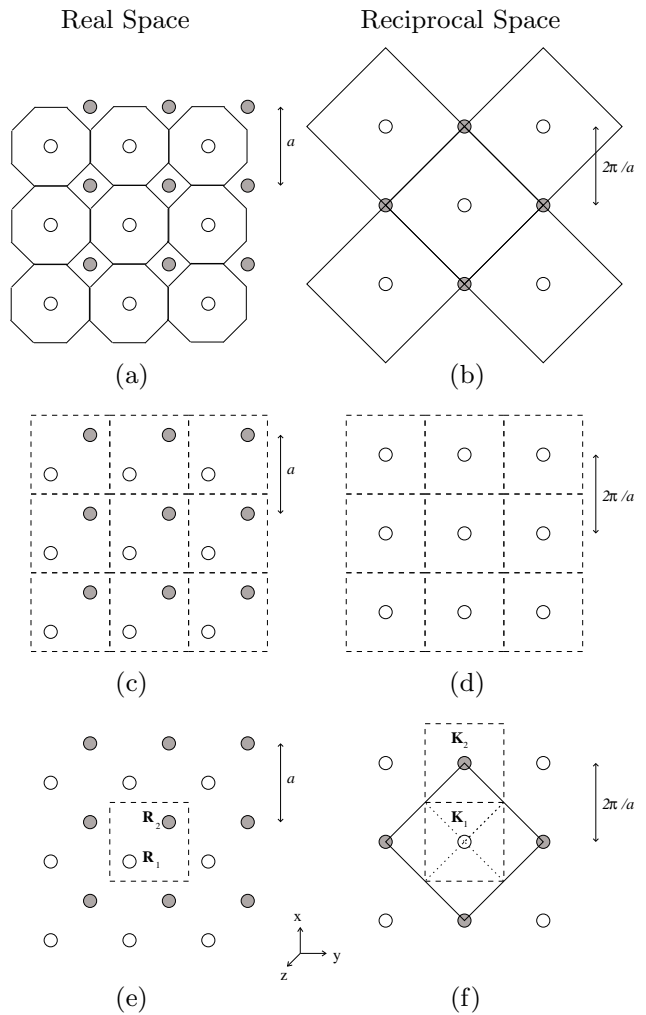}}
 \caption{\label{fig1} Example ($bcc$ lattice): (a) Cross-section of the conventional real-space tiling with Wigner-Seitz cells. The shaded sites lie out of the page. (b)  
 Cross-section of the conventional reciprocal-space tiling with Brillouin zones. The shaded sites lie out of the page. (c) Cross-section of the real-space tiling (dashed lines)
 with $N_c=2$. (d) Cross-section of the reciprocal lattice points and reciprocal-space tiles (dashed lines) of the coarse-grained system. (e) Cross-section of a real-space tile
 (dashed line) for a $N_c=2$ cluster containing the points $\b{R}_1=(0,0,0)$ and $\b{R}_2=(a/2,a/2,a/2)$. The shaded sites lie out of the page. (f) Cross-section of the
 corresponding reciprocal-space tiles (dashed lines) for the $N_c=2$ cluster, with $\b{K}_1=(0,0,0)$ and $\b{K}_2=(2\pi/a,0,0)$ at their centers. The shaded points lie out of
 the page and the solid line denotes a cross-section of the first BZ in the $(k_x,k_y)$ plane. The BZ can be visualised as a cube with a pyramid attached to each of the six
 faces, and the dotted line shows a projection of such a pyramid into the $k_z$ plane.}
 \end{center}
\end{figure}

The first step is to define the scattering path matrix $\widehat{\u{\tau}}\,^{ij}$ describing the motion of an electron in an effective medium, which ideally should be
determined so that it would describe the average properties of an electron exactly. It is a quantity which describes the full effects of the coherent potential and is given by
\begin{equation}\label{tauij}
    \label{1} \widehat{\u{\tau}}\,^{ij}=\widehat{\u{t}}\,\delta_{ij}+\sum_{k\neq i}\widehat{\u{t}}\,\left(\u{G}(\b{R}_{ik})+\widehat{\u{\delta{G}}}(\b{R}_{ik})\right)
    \,\widehat{\u{\tau}}\,^{kj}.
\end{equation}
Here a circumflex symbol denotes an effective medium quantity and an underscore denotes a matrix in angular momentum space. In addition to effective local t-matrices
$\widehat{\u{t}}$ and the usual free-space KKR structure constants $\u{G}(\b{R}_{ij})$ which account for the lattice structure, we also have effective structure constant
corrections $\widehat{\u{\delta{G}}}(\b{R}_{ij})$ which take into acccount all nonlocal scattering correlations due to the disorder configurations (labelled
$\u{\hat{\alpha}}\,^{ij}$ in Ref.~\onlinecite{Rowlands2}). Since the effective medium is translationally-invariant, the matrix elements $\widehat{\u{\tau}}\,^{ij}$ are also 
given by the Brillouin zone integral
\begin{equation}\label{tauk}
    \widehat{\u{\tau}}\,^{ij}=\frac{1}{\Omega_{BZ}}\int_{\Omega_{BZ}}\!\!\!\!d\b{k}\left(\widehat{\u{t}}^{-1}-\u{G}(\b{k})-\widehat{\u{\delta G}}(\b{k})\right)^{-1}
    \!\!e^{i\b{k}(\b{R}_i-\b{R}_j)}.
\end{equation}
Since it is not feasible to solve the problem exactly, the key idea, based upon concepts from the Dynamical Cluster Approximation, \cite{Hettler1,Jarrell1} is to perform a
consistent coarse-graining in both real and reciprocal space in order to appropriately deal with $\widehat{\u{\delta{G}}}(\b{R}_{ij})$ and $\widehat{\u{\delta{G}}}(\b{k})$
respectively. The construction for carrying out this coarse-graining, which must retain the translational invariance and point-group symmetry of the underlying lattice, has
been given by Jarrell and Krishnamurthy \cite{Jarrell1} for a 2D square lattice in connection with a simple tight-binding model Hamiltonian. We have generalised this 
construction for realistic 3D body-centered cubic ($bcc$), face-centered cubic ($fcc$), and simple cubic ($sc$) lattices \cite{Rowlands1,Rowlands2,Rowlands3} which we 
implement for the first time here.

\begin{figure}[h] 
\begin{center} 
\scalebox{0.8}{\includegraphics{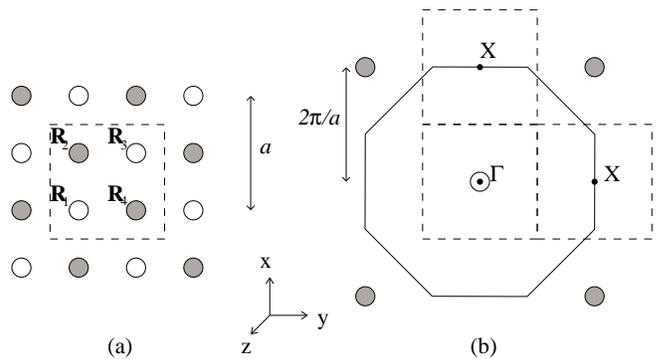}} 
 \caption{\label{fig2} (a) Cross-section of a real-space tile (dashed line) for a $N_c=4$ cluster on the $fcc$ lattice containing the points $\b{R}_1=(0,0,0)$,
 $\b{R}_2=(a/2,0,a/2)$, $\b{R}_3=(a/2,a/2,0)$ and $\b{R}_4=(0,a/2,a/2)$. The shaded sites lie out of the page. (b) Cross-section of the corresponding reciprocal-space tiles
 (dashed lines) for the $N_c=4$ cluster, with $\b{K}_1=(0,0,0)$, $\b{K}_2=(2\pi/a,0,0)$, and $\b{K}_3=(0,2\pi/a,0)$ shown as the $\Gamma$ point and the two $X$ points. The
 fourth tile is centered at the $X$ point $\b{K}_4=(0,0,2\pi/a)$ and is situated out of the page vertically above $\Gamma$. Again the shaded points lie out of the page and the
 solid line denotes a cross-section of the first BZ in the $(k_x,k_y)$ plane.} 
 \end{center}
\end{figure}

First we summarise the construction for a general lattice. Technically, the task is to find an appropriate set of $N_c$ real-space cluster sites $\{I,J\}$ and corresponding set
of `cluster momenta' $\{\b{K}_n\}$ satisfying the relation
\begin{equation} \label{IJK}
    \frac{1}{N_c}\sum_{\b{K}_n}e^{i\b{K}_n(\b{R}_I-\b{R}_J)}=\delta_{IJ}.
\end{equation}
This may be accomplished as follows:

\begin{itemize}
 
\item Choose a real-space cluster of $N_c$ sites which can be surrounded by a \emph{tile} which a) preserves the point-group symmetry of the underlying lattice and b) can be
periodically repeated to fill out all space. For $N_c=1$, the tiles are the conventional Wigner-Seitz cells surrounding each lattice point, as shown for the $bcc$ lattice in
Fig.~\ref{fig1}(a). For $N_c>1$ there may only be solutions to the problem for particular values of $N_c$ for any given lattice. For the $bcc$ lattice the next allowed cluster
sizes are $N_c=2$ and $N_c=16$, where the tiles are simple cubes of volume $a^3$ and $(2a)^3$ respectively surrounding each cluster (see Fig.~\ref{fig1}(c) for $N_c=2$).

\item Label the sites of the original lattice by the set of vectors $\{\b{R}_i^{orig}\}$, and the centers of the coarse-graining tiles by the set of vectors 
$\{\b{R}_i^{cg}\}$.

\item Label the reciprocal lattice corresponding to $\{\b{R}_i^{orig}\}$ by the set of vectors $\{\b{K}_i^{orig}\}$. Each $\b{K}_i^{orig}$ is centered in a Brillouin zone
$\Omega_{BZ}$ which periodically repeats to fill out all of reciprocal space. For the $bcc$ lattice, this Brillouin zone will be a $fcc$ Wigner-Seitz cell of volume 
$2(2{\pi}/a)^3$, as shown in Fig.~\ref{fig1}(b).

\item Label the reciprocal lattice corresponding to $\{\b{R}_i^{cg}\}$ by the set of vectors $\{\b{K}_i^{cg}\}$. Each $\b{K}_i^{cg}$ is centered at a reciprocal space tile
(corresponding to the reciprocal space of the real space tile) which again periodically repeats to fill out all of reciprocal space. For the $bcc$ lattice, $\{\b{K}_i^{cg}\}$
are simple cubic and will be centered at simple cubic tiles of volume $(2{\pi}/a)^3$ or $({\pi}/a)^3$ for the $N_c=2$ (see Fig.~\ref{fig1}(d)) or $N_c=16$ cases respectively.

\item Observe that $\{\b{K}_i^{orig}\}\subset\{\b{K}_i^{cg}\}$. Select $N_c$ vectors from the set $\{\b{K}_i^{cg}\}$ which lie within $\Omega_{BZ}$ and do not differ by an 
element of $\{\b{K}_i^{orig}\}$. We define these to be the set of cluster momenta $\{\b{K}_n\}$. The reciprocal space contained within the $N_c$ reciprocal space tiles centered 
at $\{\b{K}_n\}$ is completely equivalent to that contained within $\Omega_{BZ}$ by translation through reciprocal lattice vectors $\{\b{K}_i^{orig}\}$. See Fig.~\ref{fig1}(f) 
for the $bcc$ example with $N_c=2$. 

\end{itemize}
Refer to the table in Ref.~\onlinecite{Rowlands2} for the $\b{R}_I$ and $\b{K}_n$ values for $sc$, $bcc$ and $fcc$ lattices obtained using the above method. Note that for the 
$fcc$ lattice the next allowed cluster sizes are $N_c=4$ and $N_c=32$ respectively, where the real space tiles are again simple cubes of volume $a^3$ and $(2a)^3$ respectively 
surrounding each cluster. See Fig.~\ref{fig2} for the $fcc$ $N_c=4$ diagram.

\begin{figure}[!]
 \begin{center}
 \begin{tabular}{c}
 \scalebox{0.7}{\includegraphics{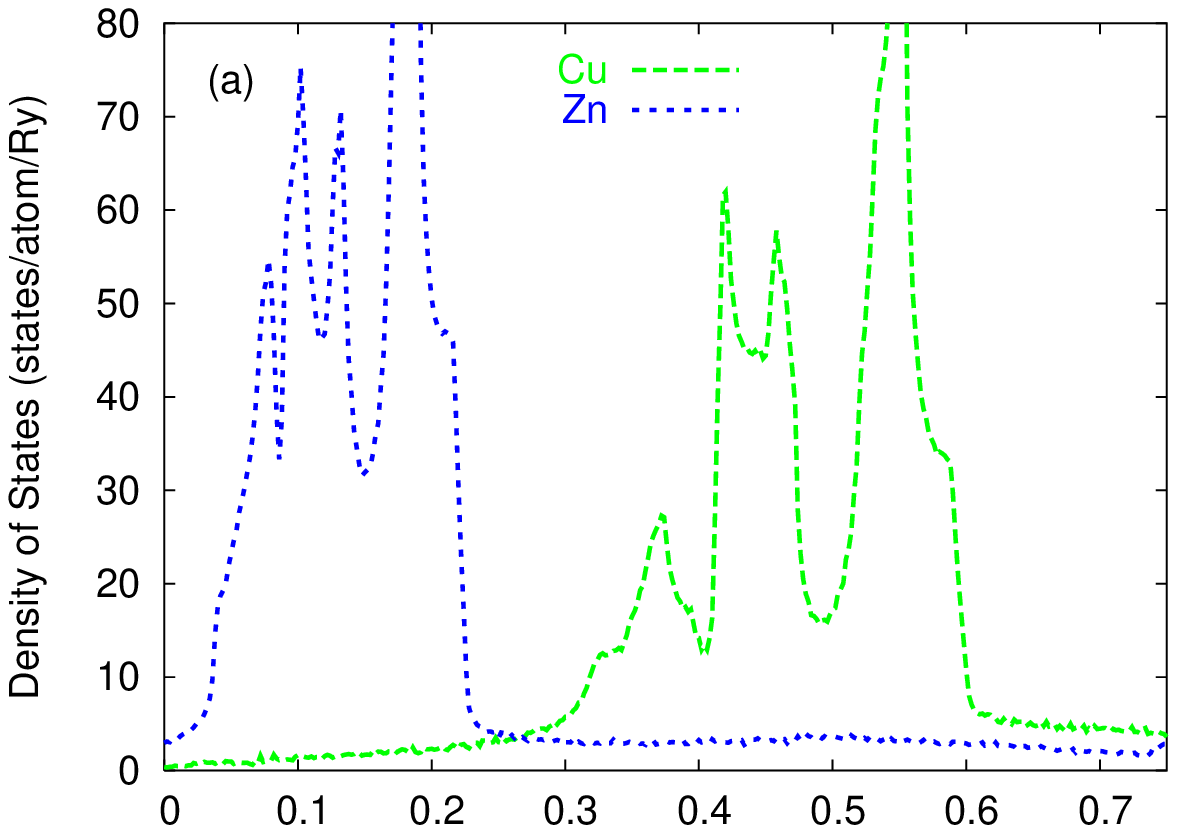}} \\
 \scalebox{0.7}{\includegraphics{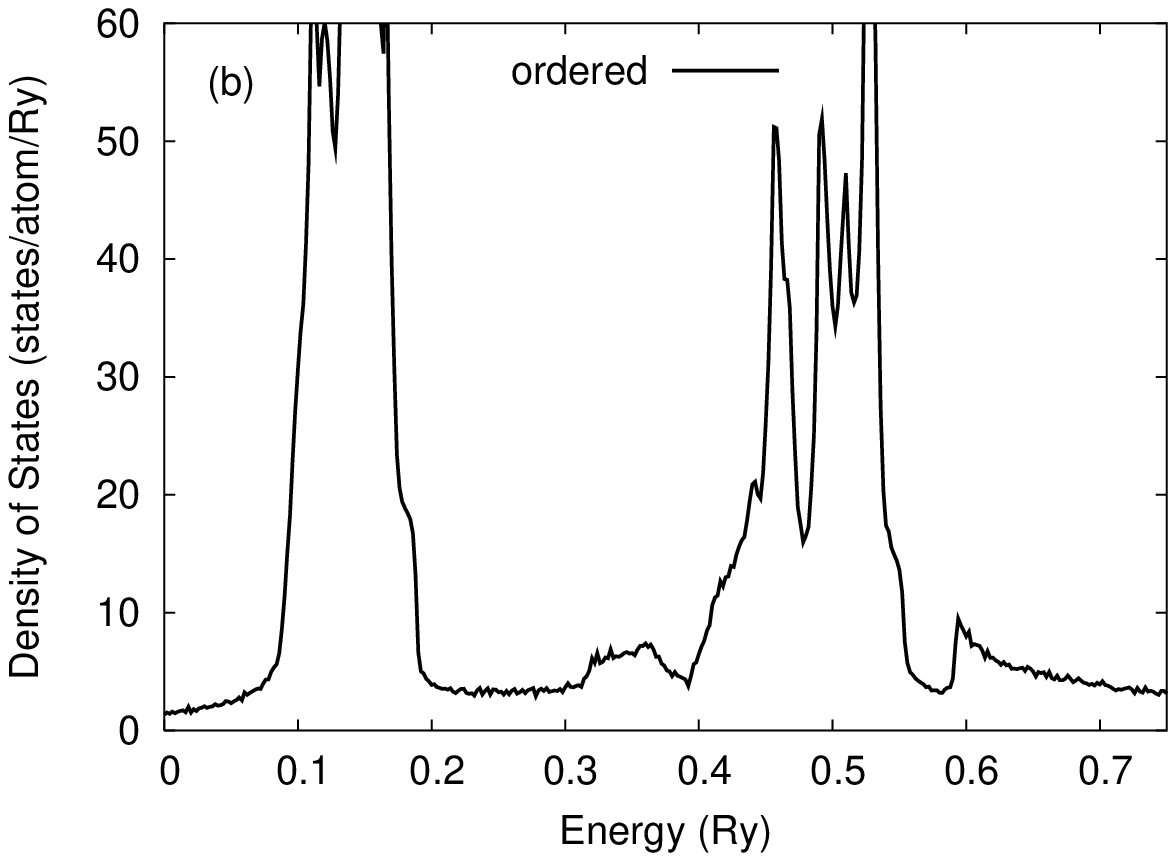}}     
 \end{tabular}   
 \caption{\label{fig3}(Color online) (a) Density of states (DOS) for pure $Cu$ and pure $Zn$. (b) DOS for ordered $Cu_{50}Zn_{50}$.}  
 \end{center}
\end{figure}

Having carried out this coarse-graining procedure, we can now make an appropriate approximation to determine the effective medium. In reciprocal space, we approximate
$\widehat{\u{\delta G}}(\b{k})$ within each of the $N_c$ tiles by the $N_c$ values $\{\widehat{\u{\delta G}}(\b{K}_n)\}$, each defined to be the average of 
$\widehat{\u{\delta G}}(\b{k})$ over the tile centered at $\b{K}_n$. The scattering path matrix may then be represented by the set of coarse-grained values
\begin{equation} \label{tauKn}
    \widehat{\u{\tau}}(\b{K}_n)=\frac{N_c}{\Omega_{BZ}}\int_{\Omega_{\b{K}_n}}\!\!\!\!d\b{k}\left(\widehat{\,\u{t}}\,^{-1}
    -\u{G}(\b{k})-\widehat{\u{\delta G}}(\b{K}_n)\right)^{-1},
\end{equation}
which are straightforward to calculate owing to \,$\widehat{\u{\delta G}}(\b{K}_n)$\, being constant within each tile \,$\Omega_{\b{K}_n}$\,. Note that the $N_c$ integrals 
here have the same computational cost as one standard BZ integral. This is unlike a supercell method such as the MCPA where the size of the KKR-matrix, which must be inverted 
at every $\b{k}$-point, increases as $N_c$ increases. In fact it is straightforward to show that this integration step is $N_c$ times faster for the KKR-NLCPA than for a
supercell method for comparable cluster sizes. Using Eq.~\ref{IJK}, the scattering path matrix at the cluster sites becomes
\begin{eqnarray} \label{tauIJK}
    \widehat{\u{\tau}}\ ^{IJ}\!=\frac{1}{\Omega_{BZ}}\sum_{\b{K}_n}\left(\int_{\Omega_{\b{K}_n}}\!\!\!\!\!d\b{k}
    \left(\,\widehat{\u{t}}\,^{-1}-\u{G}(\b{k})-\widehat{\u{\delta G}}(\b{K}_n)\right)^{-1}\right) & & \nonumber \\
\times e^{i\b{K}_n(\b{R}_I-\b{R}_J)} \ \ \ \ & &
\end{eqnarray}
From Nyquist's sampling theorem, \cite{Elliot1,Jarrell1} the effect of coarse-graining the effective structure constant corrections is to reduce their range in real space. In 
fact from Eq.~\ref{IJK} we have
\begin{eqnarray}\label{IJtoK}
    \widehat{\u{\delta{G}}}(\b{R}_{IJ})&=&\frac{1}{N_c}\sum_{\b{K}_n}\widehat{\u{\delta G}}(\b{K}_n)e^{i\b{K}_n(\b{R}_I-\b{R}_J)} \nonumber \\
    \widehat{\u{\delta G}}(\b{K}_n)&=&\sum_{J\neq I}\widehat{\u{\delta{G}}}(\b{R}_{IJ})e^{-i\b{K}_n(\b{R}_I-\b{R}_J)}.
\end{eqnarray}
Note that $\widehat{\u{\delta{G}}}(\b{R}_{IJ})$ remains a translationally-invariant quantity which depends only on the distance between sites $I$ and $J$, now within the range
of the cluster size, but independent of which site in the lattice is chosen to be site $I$. \cite{Rowlands1} It is now straightforward to generalise the CPA argument and
determine the medium by mapping to an impurity cluster problem. We choose a real-space cluster consistent with the requirements outlined above, and use the Embedded Cluster
Method \cite{Gonis2,Gonis1} to replace it with an `impurity' cluster of real t-matrices and free-space structure constants in the (still undetermined) effective medium. We then
consider all paths starting and ending on the impurity cluster sites and demand that the average over the $2^{N_c}$ possible impurity cluster configurations $\gamma$ be equal
to the path matrix for the effective medium itself i.e.
\begin{equation} \label{nlcpacond}
    \sum_{\gamma}P(\gamma)\,\u{\tau}_{\gamma}^{\,IJ}=\widehat{\u{\tau}}\,^{IJ},
\end{equation}
where $P(\gamma)$ is the probability of configuration $\gamma$ occuring. Therefore the effective medium t-matrices and effective structure constants are determined from a
self-consistent solution of Eqs.~\ref{tauIJK} and \ref{nlcpacond}. An example algorithm is given in Refs.~\onlinecite{Rowlands1,Rowlands2}. SRO may be included by appropriately
weighting the configurations in Eq.~\ref{nlcpacond} (the number of which can be reduced using symmetry and sampling), provided that translational-invariance is preserved.
Observable quantities such as the density of states (DOS) can be calculated from the corresponding configurationally-averaged Green's function. \cite{Faulkner1} The formula for
the DOS within the KKR-NLCPA is given in Refs.~\onlinecite{Rowlands1,Rowlands2}. Importantly owing to the translational invariance of the KKR-NLCPA medium it is independent of
lattice site chosen. This is crucial for calculating the partially-averaged charge densities to be used in combination with DFT. 
\cite{Rowlands1,Rowlands4,Bruno2,Faulkner3,Faulkner4} Finally, note that the KKR-NLCPA formalism reduces to the KKR-CPA for $N_c=1$, and nonlocal scattering correlations (and
SRO if desired) are systematically included into the effective medium as $N_c$ is increased, becoming exact as $N_c\rightarrow\infty$.

\section{Results}

\begin{figure}[!]
 \begin{center}
 \begin{tabular}{c}
 \scalebox{0.7}{\includegraphics{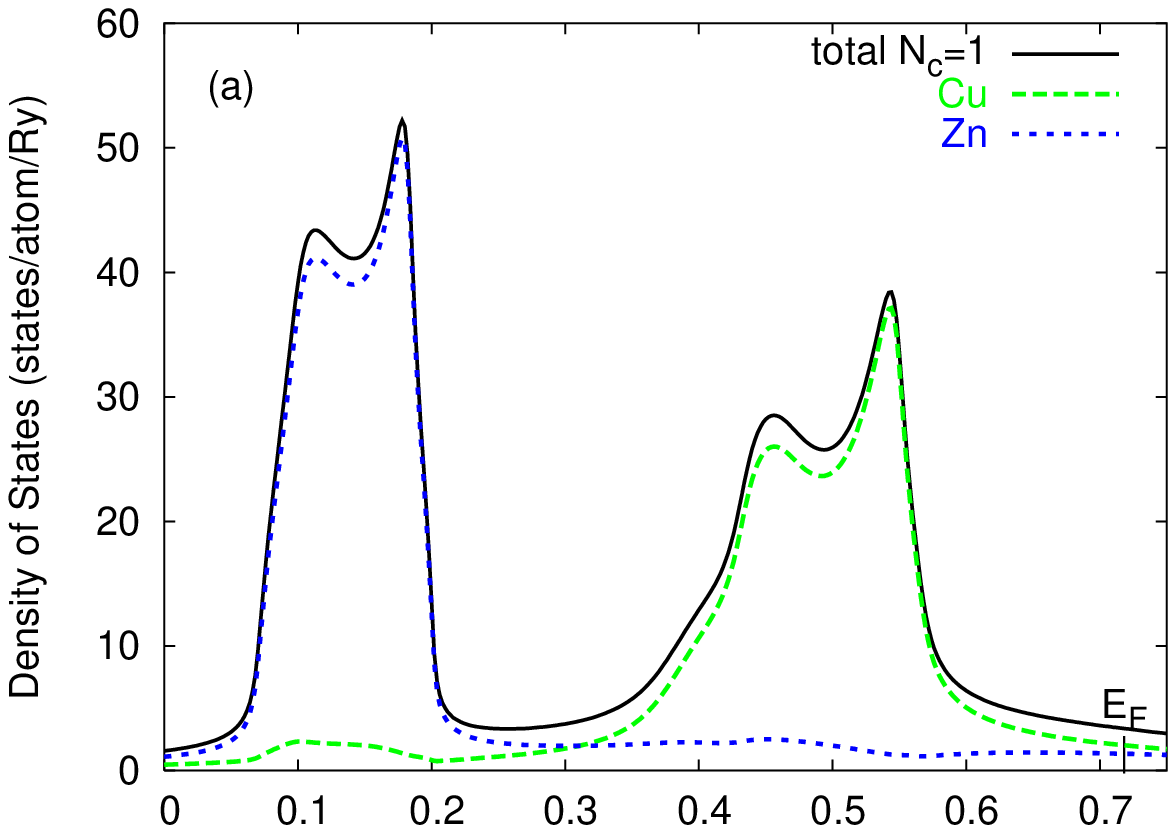}} \\ 
 \scalebox{0.7}{\includegraphics{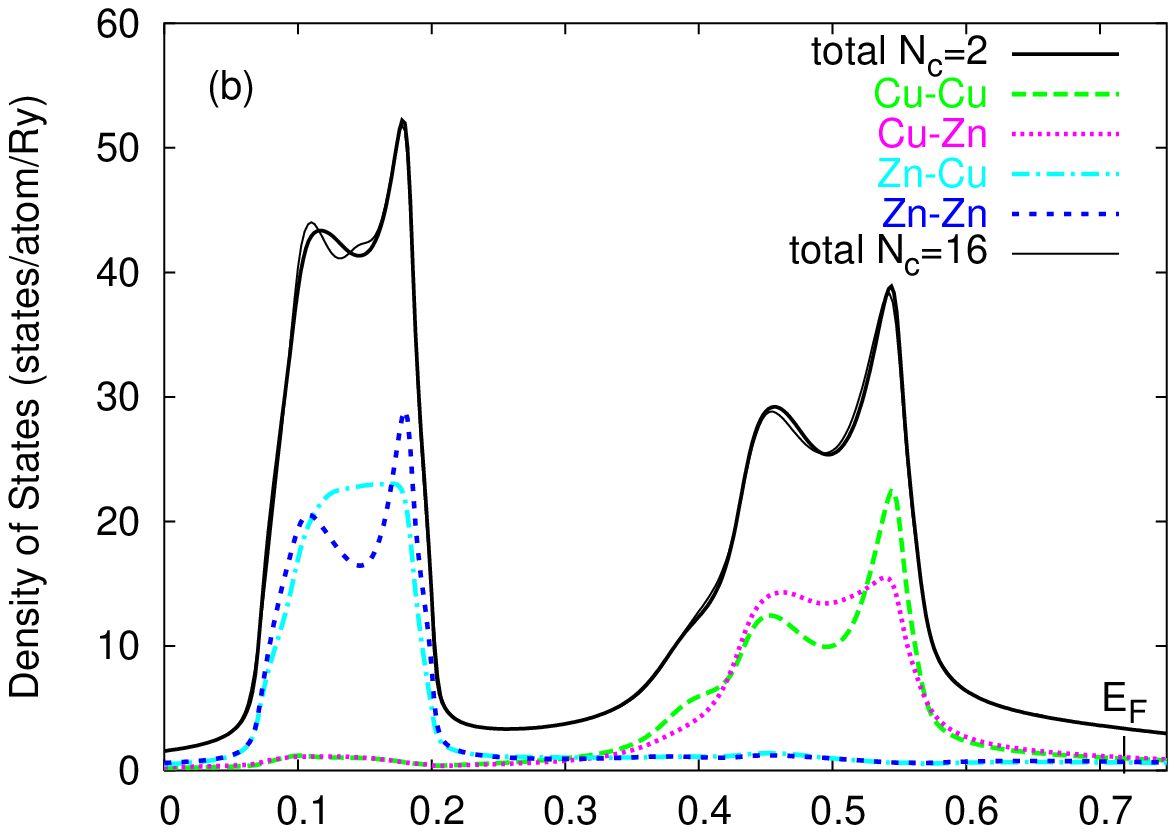}} \\
 \scalebox{0.7}{\includegraphics{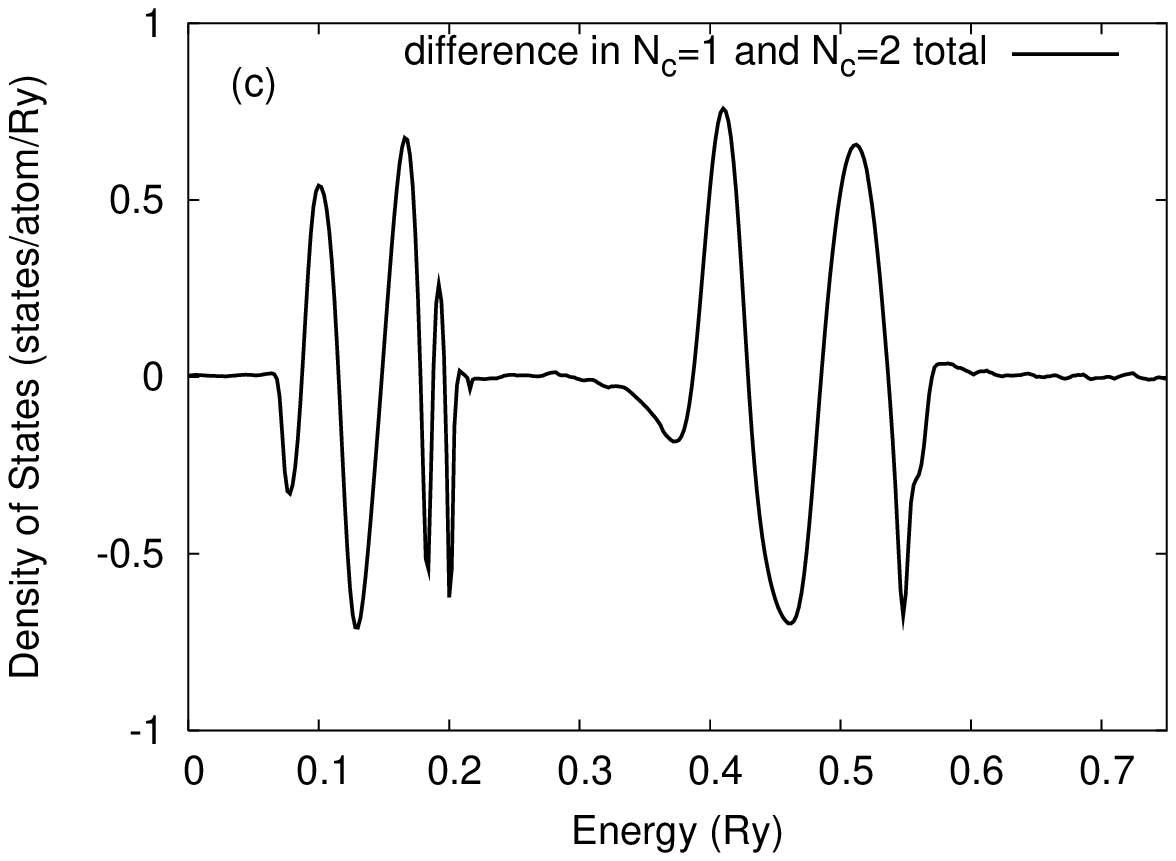}} 
 \end{tabular} 
 \caption{\label{fig4}(Color online) (a) Total average density of states (DOS) for disordered $bcc$ $Cu_{50}Zn_{50}$ using the KKR-CPA. Also shown are the contributions from 
 the $Cu$ and $Zn$ components (site restricted average DOS). $E_F$ is the Fermi energy. (b) Total average DOS for $bcc$ $Cu_{50}Zn_{50}$ using the KKR-NLCPA with $N_c=2$, along 
 with the contributions from the 4 possible cluster configurations (cluster restricted average DOS) measured at the first site i.e.~$Cu$ for $Cu$-$Cu$, $Cu$-$Zn$, and $Zn$ for 
 $Zn$-$Cu$, $Zn$-$Zn$. (Owing to the translational invariance, contributions measured at the second site would give the same results with a simple reversal of the labels). Also
 shown are total DOS results for $N_c=16$. (c) Plot of the difference in the total DOS between the $N_c=1$ and $N_c=2$ calculations i.e.~(total $N_c=1$)$-$(total $N_c=2$).} 
 \end{center}
\end{figure}

\begin{figure*}[!]   
 \begin{center}
 \begin{tabular}{cc}
 \scalebox{0.7}{\includegraphics{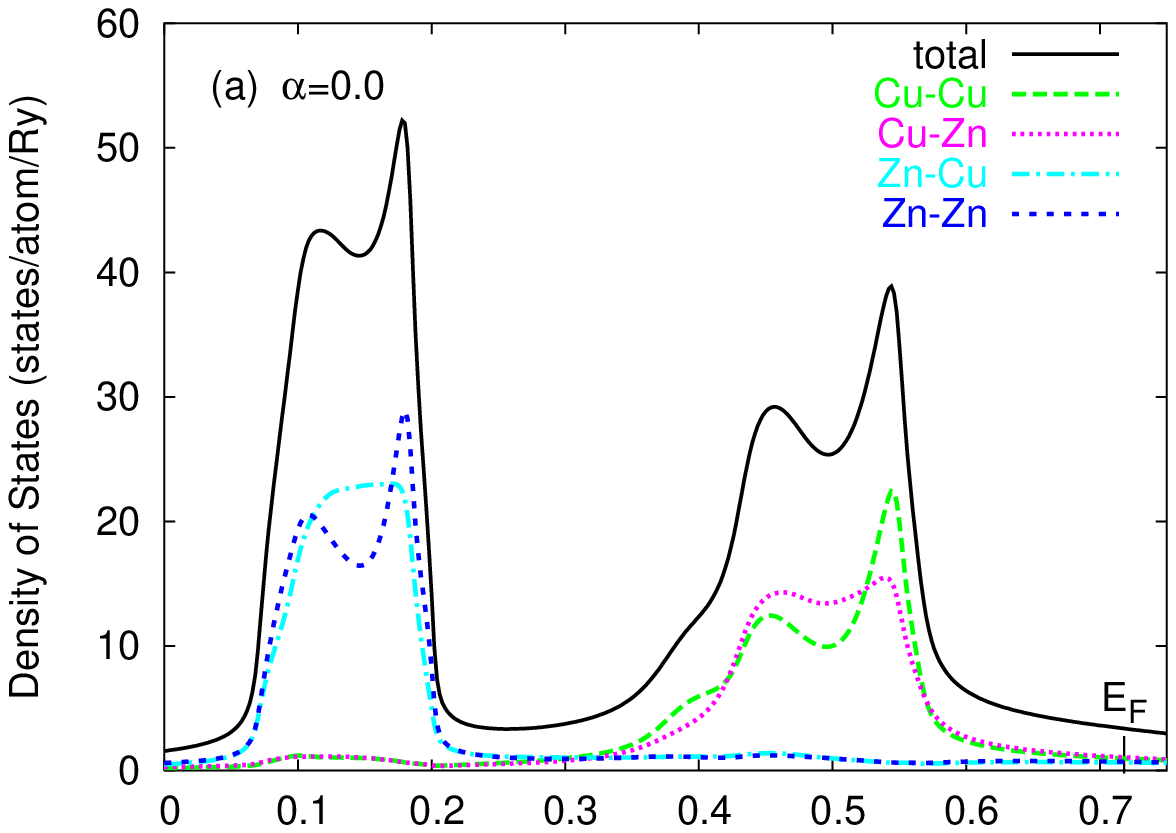}}  &  \scalebox{0.7}{\includegraphics{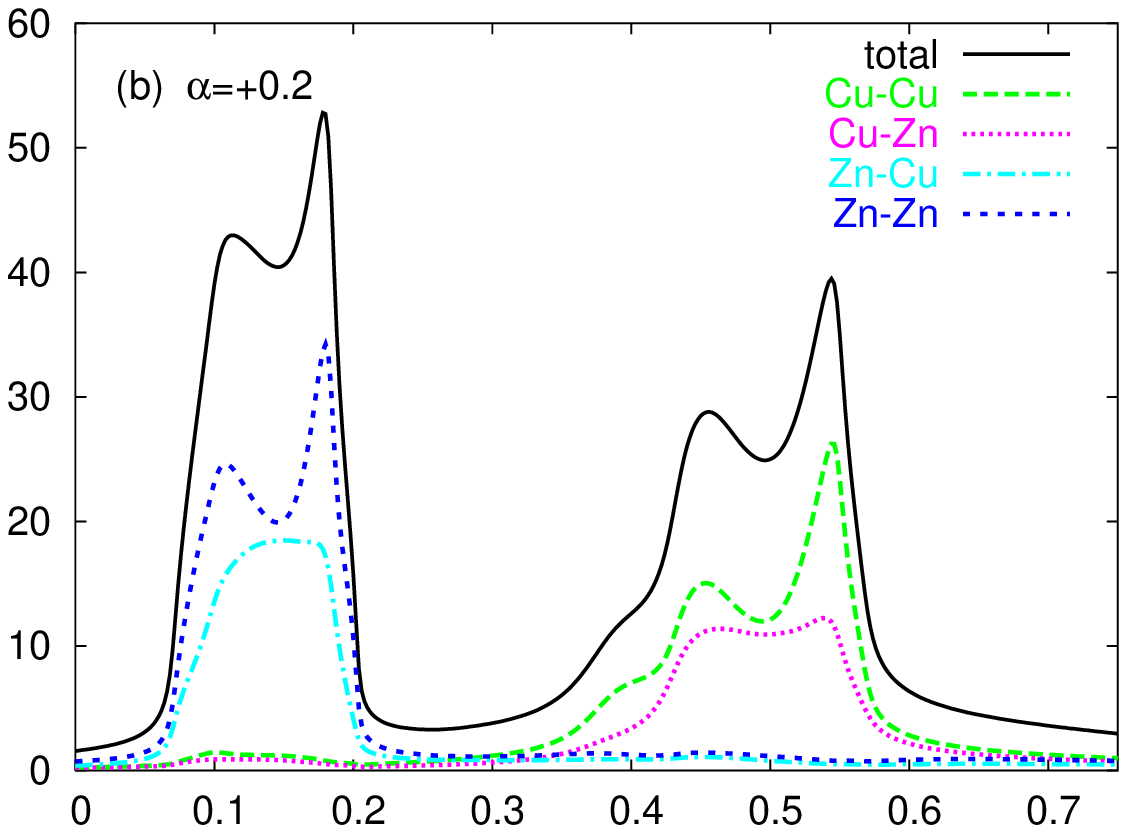}} \\
 \scalebox{0.7}{\includegraphics{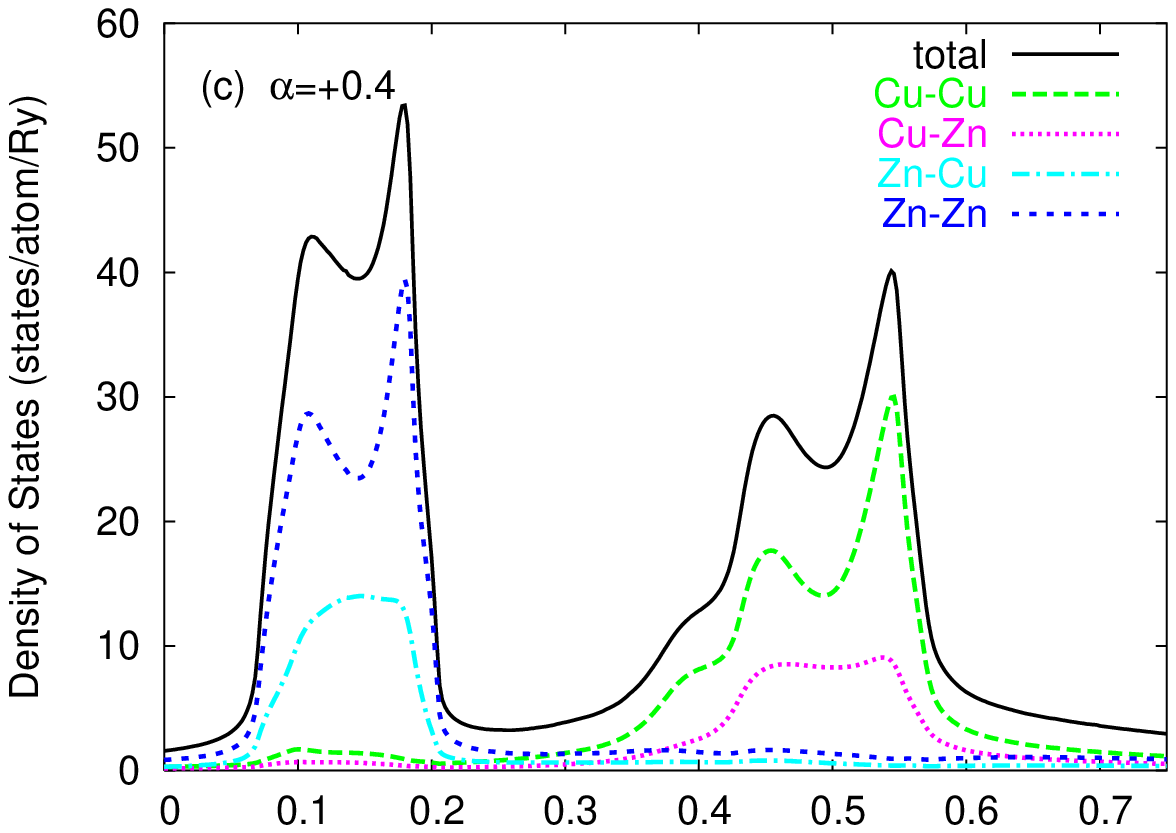}} &  \scalebox{0.7}{\includegraphics{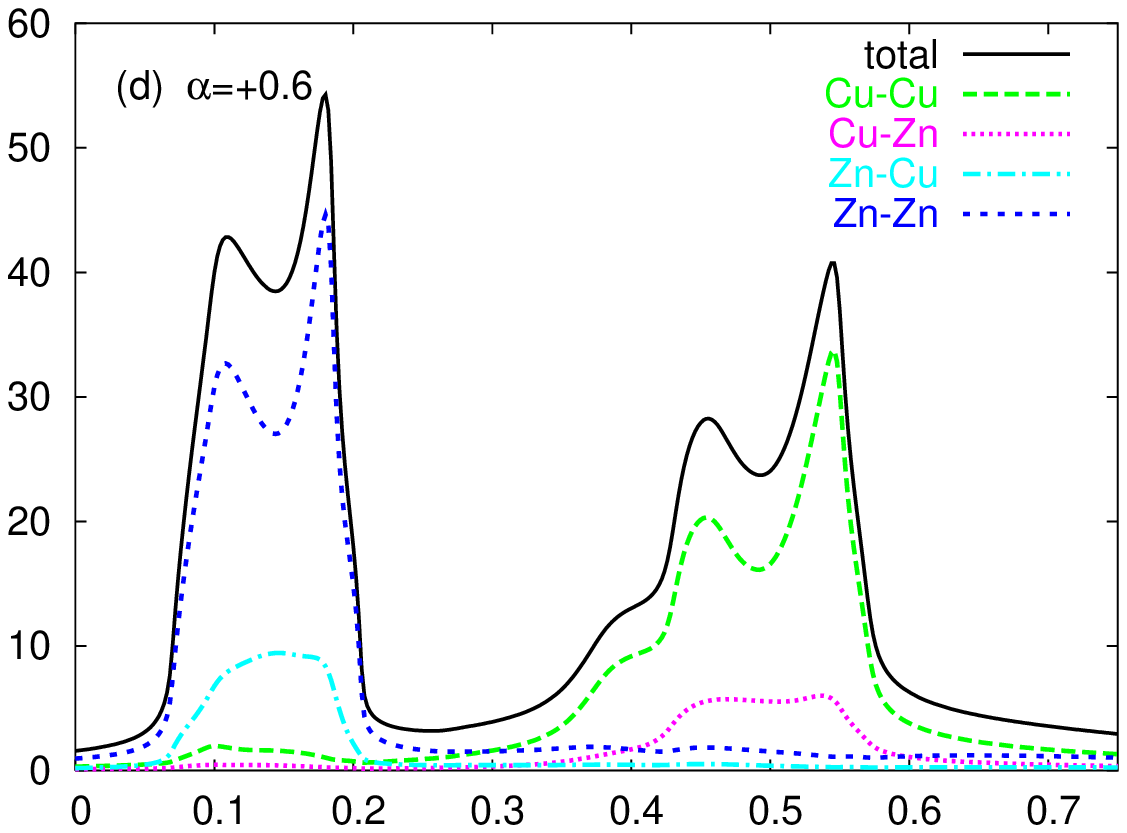}} \\ 
 \scalebox{0.7}{\includegraphics{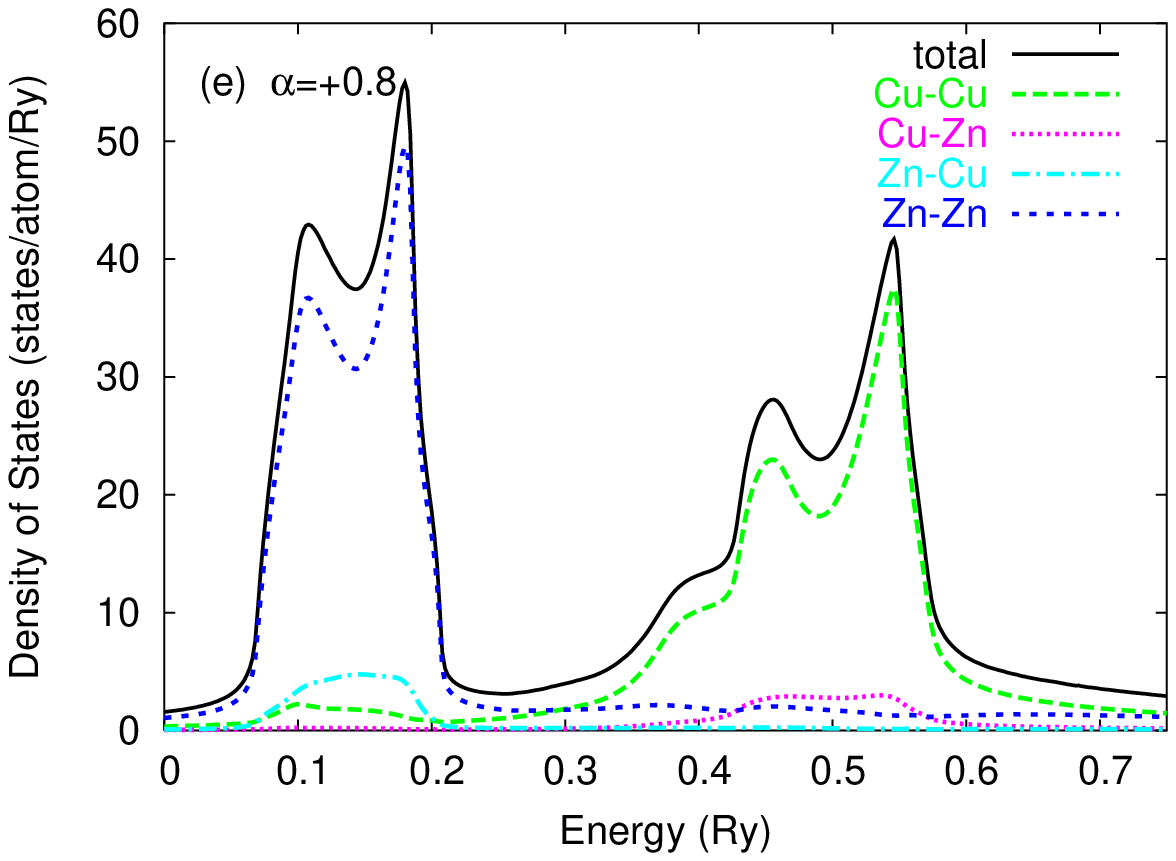}} &  \scalebox{0.7}{\includegraphics{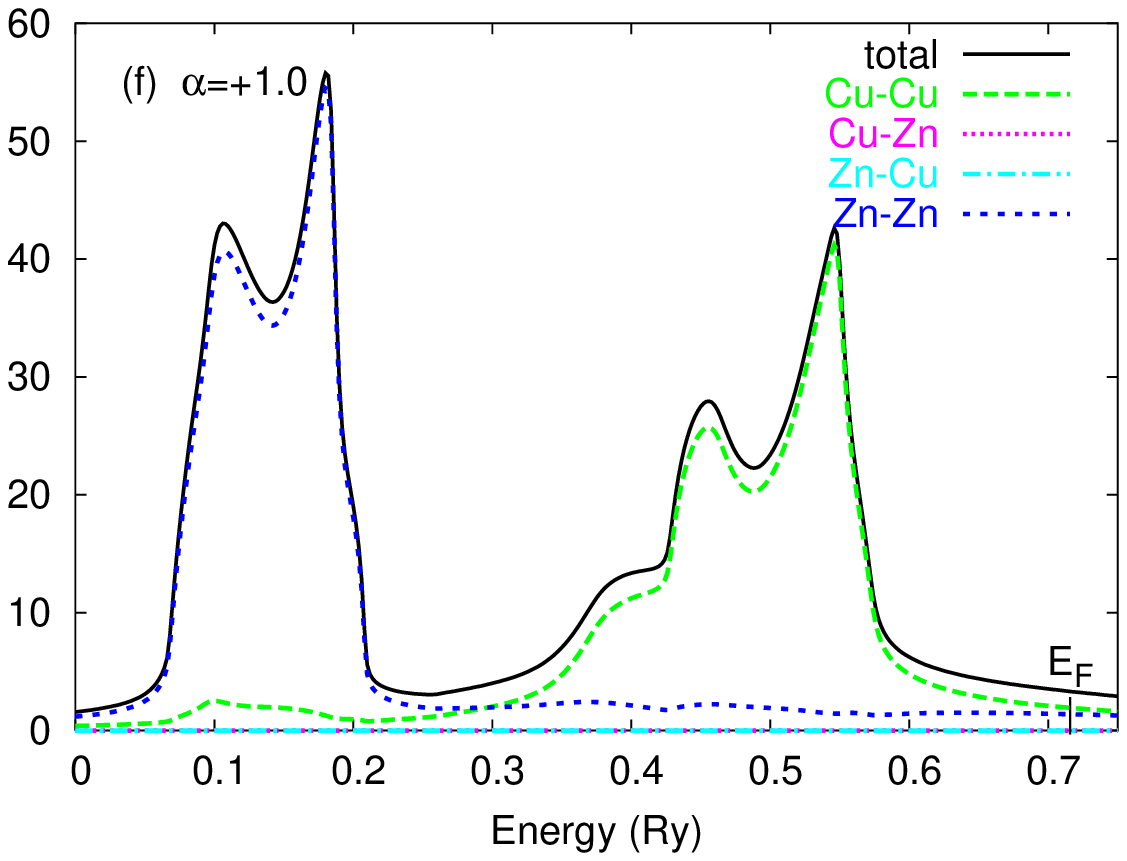}} 
 \end{tabular}          
 \caption{\label{fig5}(Color online) (a) Total and cluster component density of states for $bcc$ $Cu_{50}Zn_{50}$ using the KKR-NLCPA with $N_c=2$ and
 $\alpha=0$. (b)$\rightarrow$(f) Same as (a) but with increasing values of $\alpha$, corresponding to short-range clustering.}     
 \end{center}
\end{figure*}

\begin{figure*}[!]
 \begin{center}
 \begin{tabular}{cc}
 \scalebox{0.7}{\includegraphics{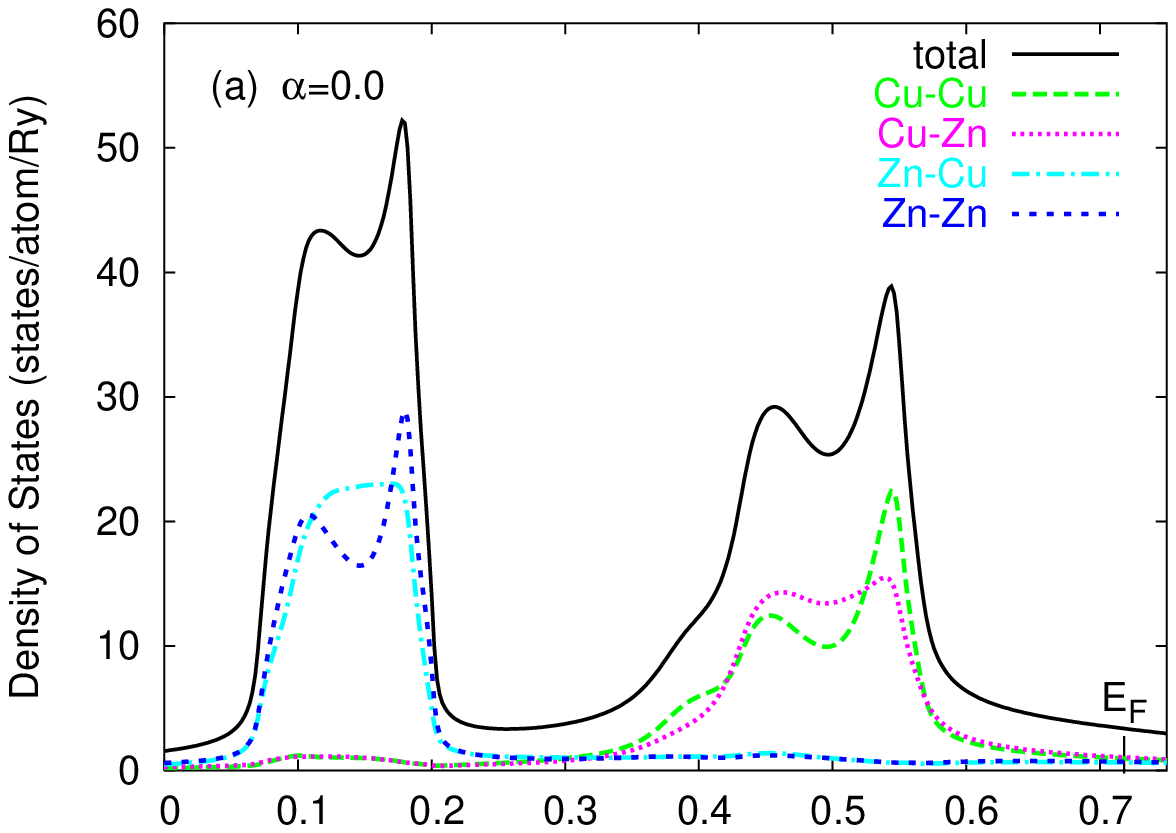}}  &  \scalebox{0.7}{\includegraphics{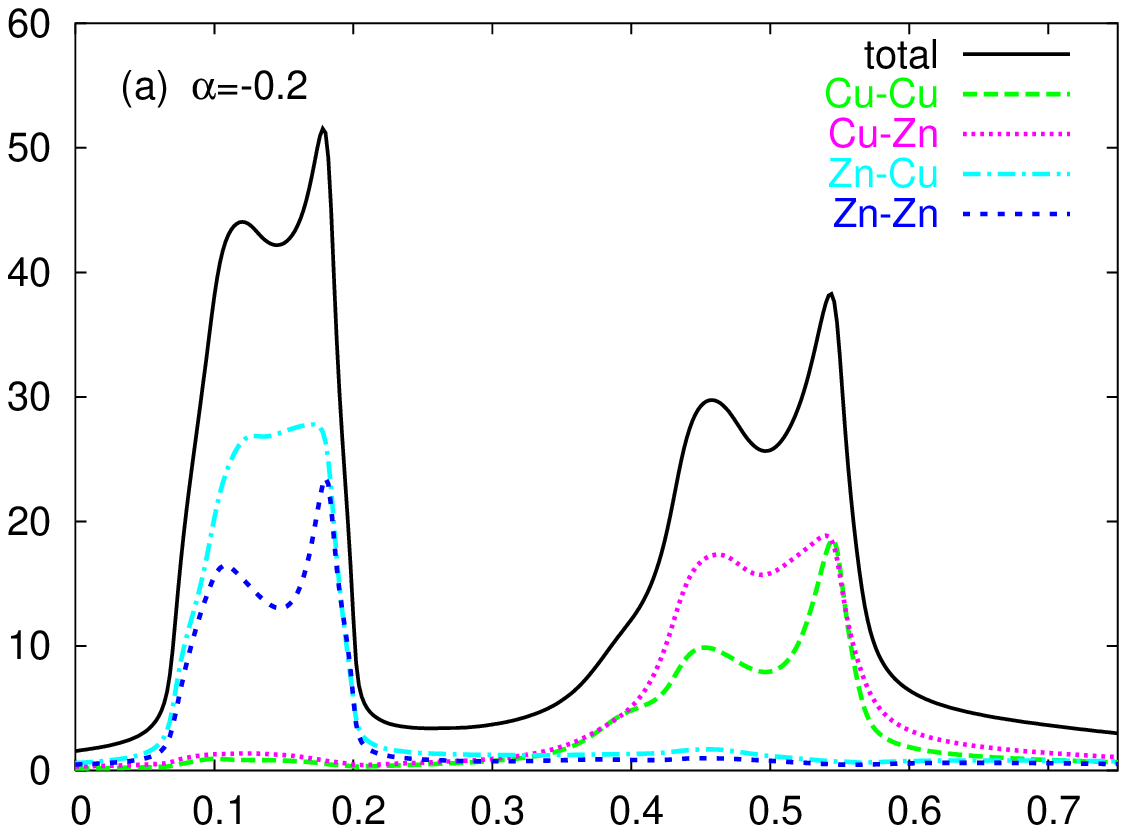}} \\
 \scalebox{0.7}{\includegraphics{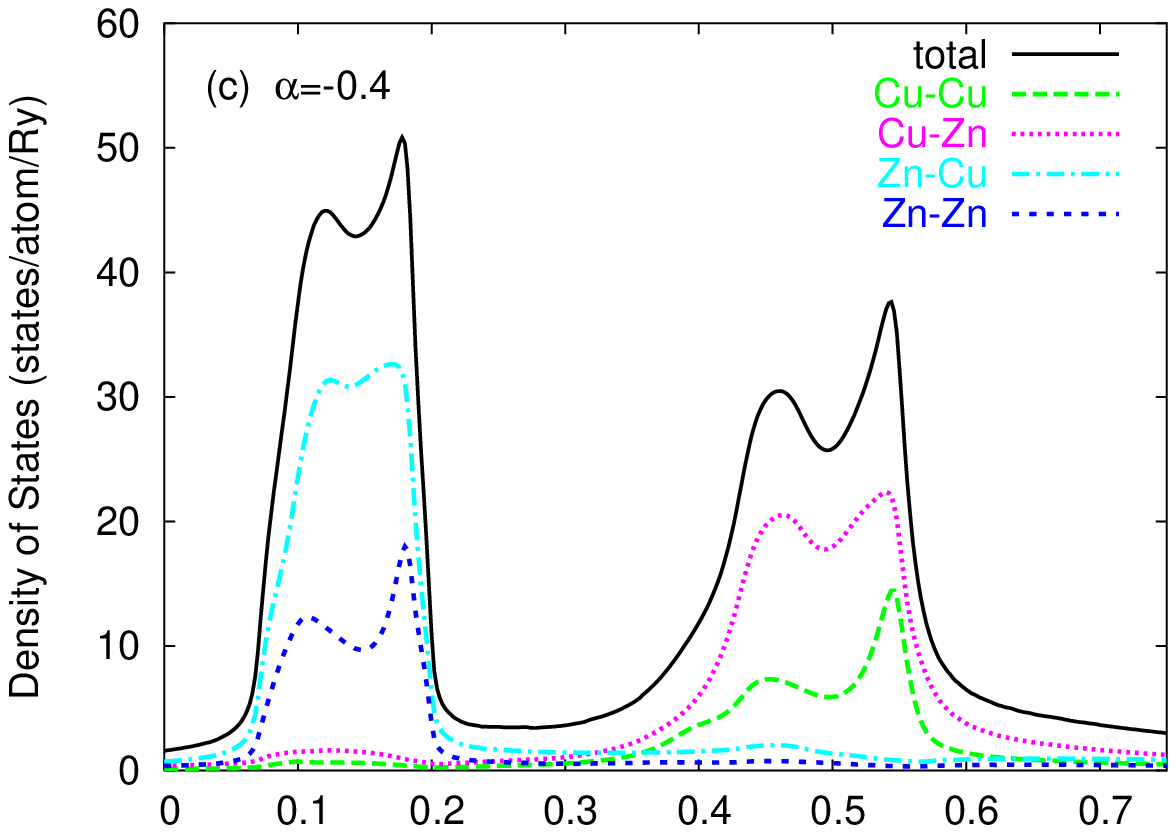}} &  \scalebox{0.7}{\includegraphics{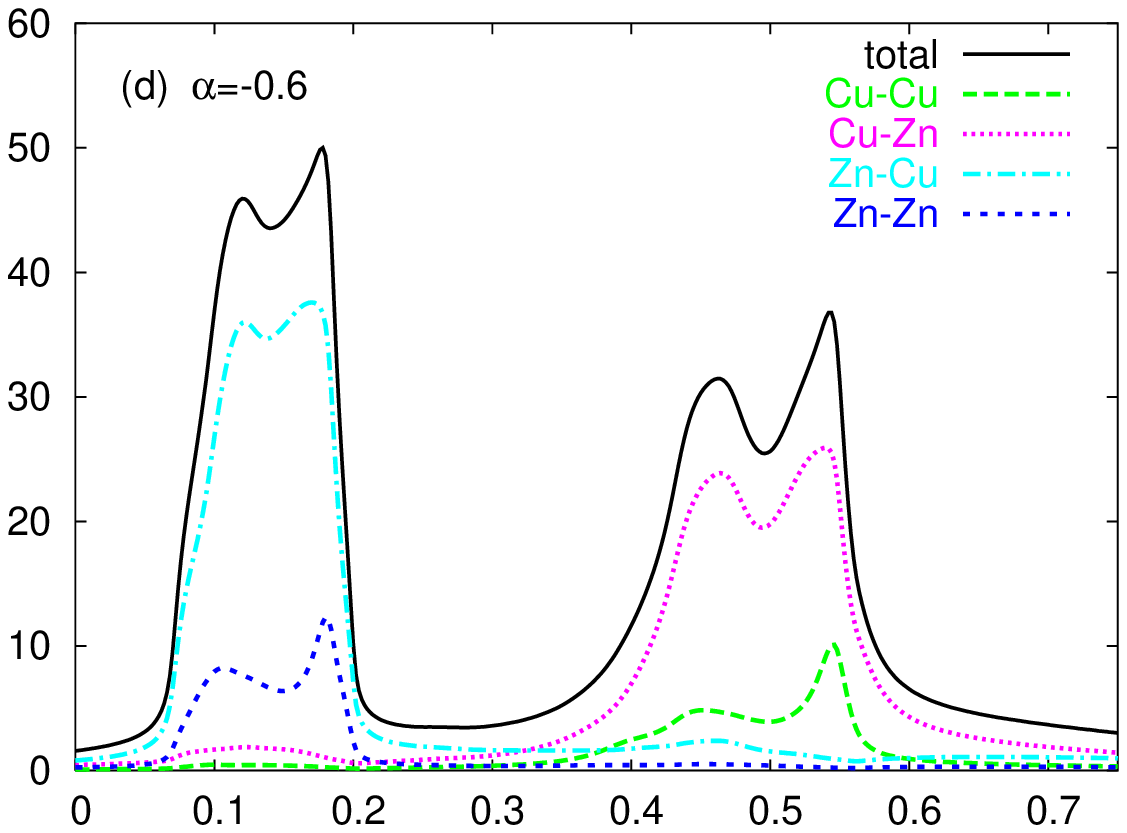}} \\ 
 \scalebox{0.7}{\includegraphics{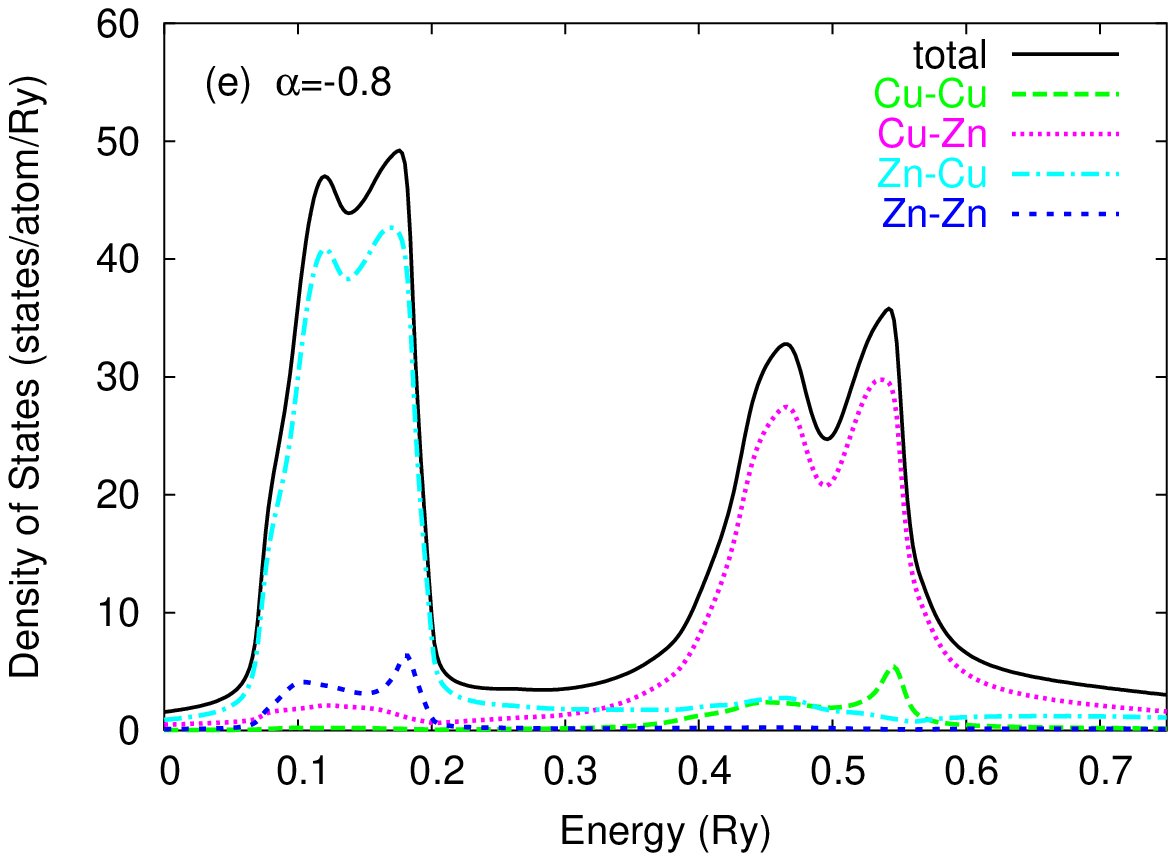}} &  \scalebox{0.7}{\includegraphics{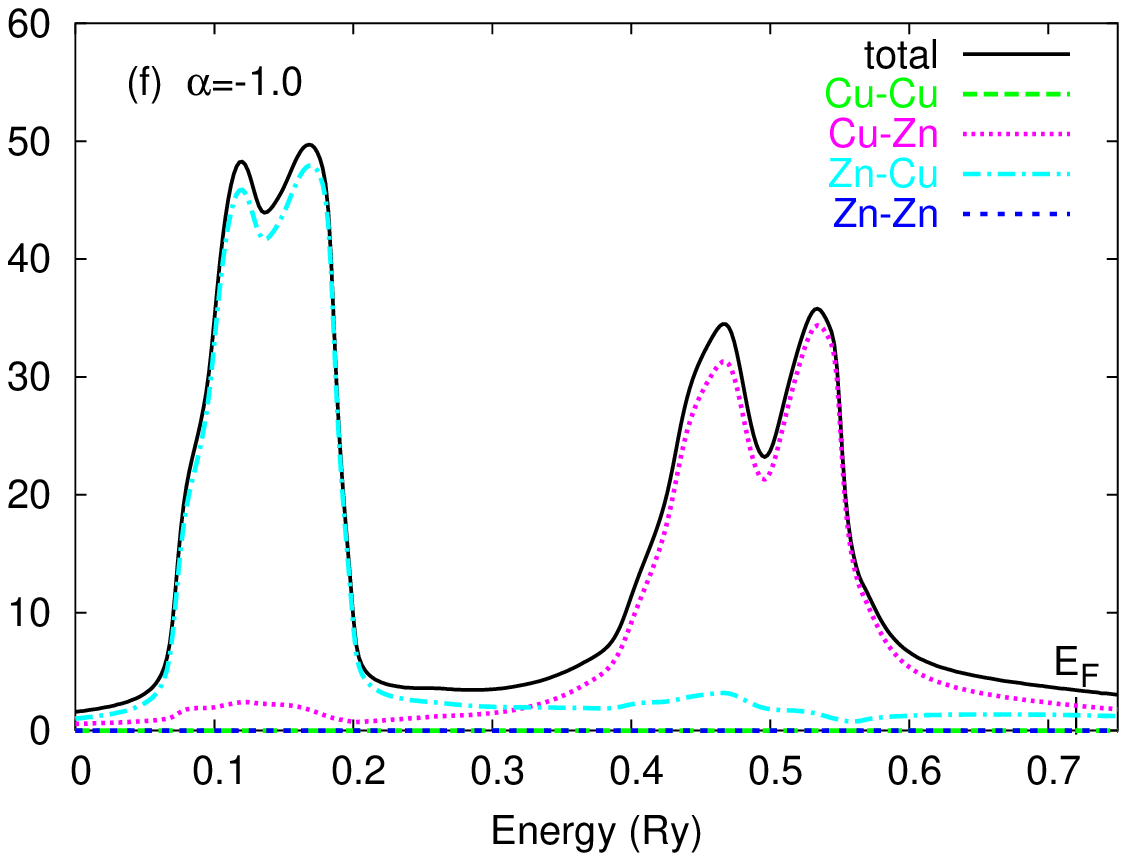}} 
 \end{tabular}         
 \caption{\label{fig6}(Color online) (a) Total and cluster component density of states for $bcc$ $Cu_{50}Zn_{50}$ using the KKR-NLCPA with $N_c=2$ and
 $\alpha=0$. (b)$\rightarrow$(f) Same as (a) but with decreasing values of $\alpha$, corresponding to short-range ordering.}
 \end{center}
\end{figure*}

To illustrate the KKR-NLCPA, we present calculations for the $bcc$ $Cu_{50}Zn_{50}$ solid solution with lattice constant 2.86\AA. In all calculations that follow, $Cu$ and $Zn$
potentials come from self-consistent field (SCF) KKR-CPA calculations, \cite{Stocks2,Johnson3} the Brillouin zone integrals use the adaptive quadrature method, \cite{Bruno1}
and the energy contour has 1~mRy imaginary part. 

Fig.~\ref{fig3}(a) shows DOS plots for pure $Cu$ and pure $Zn$, and Fig.~\ref{fig3}(b) shows a calculation for the ordered $Cu_{50}Zn_{50}$ compound. Since the energies of the
$Cu$ and $Zn$ $d$-bands are very different, the system is said to be in the `split band' regime. Physically, this means an electron travels more easily between $Cu$ or between
$Zn$ sites than between unlike sites and so the decrease in overlap between like sites in the ordered case results in a narrowing of the $Cu$ and $Zn$ bands by a factor of two
compared with the pure calculations. \cite{Moruzzi1} Fig.~\ref{fig4}(a) shows KKR-CPA results for disordered $Cu_{50}Zn_{50}$. It is clear that the bands are widened and
smoothened compared with the DOS for the ordered  calculation. The component contributions from $Cu$ and $Zn$ impurity sites embedded in the KKR-CPA medium are also shown.
Next, a KKR-NLCPA calculation for disordered $Cu_{50}Zn_{50}$ for a two-site cluster ($N_c=2$) is shown in Fig.~\ref{fig4}(b). First note that there is little observable
difference in the total DOS compared with the KKR-CPA calculation. This is due to the small size of the cluster, and the difference due to the nonlocal scattering correlations
shows up in detail only on a scale of $\pm{1}$ state/atom/Ry, as shown in Fig.~\ref{fig4}(c). As expected, it is clear that the extra structure is in the energy regions of the
impurity $d$-bands. However the most striking aspect of the KKR-NLCPA calculation is that the component contributions to the total DOS from the four possible cluster
configurations are apparent. The component plots here are the DOS measured at the first cluster site when a particular cluster configuration is embedded in the KKR-NLCPA
medium, which is the $Cu$ site for the $Cu$-$Cu$ and $Cu$-$Zn$ configurations, and the $Zn$ site for the $Zn$-$Cu$ and $Zn$-$Zn$ configurations. Crucially, owing to the
translational invariance of the KKR-NLCPA medium, measurement at the second site gives the same results with a simple reversal of the labels of the $Cu$-$Zn$ and $Zn$-$Cu$
components. These component plots are particularly useful for interpreting the effects of SRO on the electronic structure, as described in the next section. Finally,
Fig.~\ref{fig4}(b) also shows total DOS results for the larger $N_c=16$ cluster, where the extra structure is more visible in the energy region between 0.10~Ry and 0.15~Ry 
where some states are shifted to higher energies.

\subsection*{Short-Range Order}

Unlike single-site theories, it is possible to include the effects of short-range order (SRO) on the electronic structure of disordered systems using the KKR-NLCPA. This may be
done by using an appropriate non-random probability distribution when averaging over the impurity cluster configurations in Eq.~\ref{nlcpacond}. Of course as the size of the
cluster increases, the range of possible SRO that may be included also increases. Here we show results for $bcc$ $Cu_{50}Zn_{50}$ using a pair cluster ($N_c=2$), where it is
possible to include SRO between nearest neighbour sites only. This may be straightforwardly done by introducing the nearest neighbour Warren-Cowley SRO parameter $\alpha$,
\cite{Cowley1} and using probabilities defined as  
\begin{eqnarray*} 
P(CuCu) & = & P(Cu)^2+\alpha/4 \\ 
P(ZnZn) & = & P(Zn)^2+\alpha/4 \\ 
P(CuZn) & = & P(Cu)P(Zn)-\alpha/4 \\ 
P(ZnCu) & = & P(Zn)P(Cu)-\alpha/4  
\end{eqnarray*}
For $Cu_{50}Zn_{50}$, $P(Cu)=P(Zn)=0.5$ and so the SRO parameter can take values in the range ${-1}\leq\alpha\leq{1}$, where $-1$, $0$ and $+1$ correspond to ideal ordering, 
complete randomness, and ideal clustering. 

Fig.~\ref{fig5} shows the effects of short-range clustering upon the configurationally-averaged DOS. It is clear that as $\alpha$ increases above zero, the probability of like
pair components increases whilst that of unlike pairs decreases, resulting in corresponding changes to the component and total DOS. Indeed at $\alpha=+1$ the probability of
unlike pairs is zero and the total DOS is now completely dominated by the features of the $Cu$-$Cu$ and $Zn$-$Zn$ components, as shown in Fig.~\ref{fig5}(f). As expected these
features are reminiscent of the pure bands shown in Fig.~\ref{fig3}(a), for example the magnifying of the trough and peak either side of 0.15~Ry and 0.5~Ry can all be
associated with the DOS for pure $Zn$ and pure $Cu$ respectively. Moreover, a new peak appears just before 0.4~Ry associated with pure $Cu$.  Fig.~\ref{fig6} shows the effects
of short-range ordering upon the configurationally-averaged DOS. Here as $\alpha$ decreases below zero, the components of the total DOS due to like pairs  decreases whilst that
due to unlike pairs increases. Finally at $\alpha=-1$, as shown in Fig.~\ref{fig6}(f), there are only contributions remaining from unlike pairs. Evidently, the DOS in general
has a closer resemblance to that of ordered $Cu_{50}Zn_{50}$ shown in Fig.~\ref{fig3}(b). For example the peaks either side of 0.15~Ry and 0.5~Ry are now of roughly equal
magnitude and there is also a slight overall narrowing of the bands compared with those for positive values of $\alpha$ due to the decrease in probability of like neighbours.

\section{Conclusions}

In summary, the recently devised KKR-NLCPA generalises the widely used KKR-CPA method by including SRO, and satisfies all of the requirements for a satisfactory effective
medium cluster theory. \cite{Gonis1} It has relatively low computational cost in comparison with supercell-based methods since the BZ integration, the most
computationally-demanding aspect of a band structure calculation, does not scale with the cluster size. In this paper we have implemented the KKR-NLCPA for a realistic system 
by illustrating the dramatic changes that can occur in the DOS for $bcc$ $Cu_{50}Zn_{50}$.

Ultimately the KKR-NLCPA will be the electronic structure component of a fully self-consistent theory of disordered systems. The next step will be to combine it with DFT, 
\cite{Rowlands4} which will also enable charge correlations \cite{Bruno2,Faulkner3,Faulkner4} and local lattice displacements \cite{Stefanou2} to be systematically taken into
account for alloys. Then, the SRO parameter $\alpha$ will need to be coarse-grained via Eqs.~\ref{IJtoK}, and could be determined via a linear response calculation
\cite{Staunton1} before being fed back into the electronic structure, resulting in a completely ab-initio theory of SRO at a given temperature $T$. Similar treatments will also
be available for effects such as magnetic SRO in metallic magnets at finite temperature.

\begin{acknowledgments}
Thanks to S.~B.~Dugdale and S.~Ostanin for computational assistance. We acknowledge support from EPSRC~(UK).
\end{acknowledgments}


\end{document}